\documentclass[journal,10pt]{IEEEtran}

\usepackage{cite}
\usepackage{amsmath,amssymb,amsfonts}
\usepackage{algorithmic}
\usepackage{graphicx}
\usepackage{textcomp}
\usepackage{xcolor}
\def\BibTeX{{\rm B\kern-.05em{\sc i\kern-.025em b}\kern-.08em
    T\kern-.1667em\lower.7ex\hbox{E}\kern-.125emX}}
\usepackage{graphicx}
\usepackage{subcaption}
\graphicspath{ {images/} }
\usepackage[utf8]{inputenc}
\usepackage{fancyhdr}
\usepackage{dirtytalk}
\usepackage{tabularx}
\PassOptionsToPackage{hyphens}{url}\usepackage{hyperref}
\usepackage[ruled,vlined]{algorithm2e}

\usepackage{enumitem,kantlipsum}

\usepackage{graphicx}
\graphicspath{ {images/} }
\usepackage[utf8]{inputenc}
\usepackage{fancyhdr}
\usepackage{dirtytalk}
\usepackage{tikz}
\usetikzlibrary{shapes.geometric, arrows}
\usepackage{pgfplots}
\pgfplotsset{width=8cm,compat=1.16}
\usepgfplotslibrary{external}
\tikzstyle{startstop} = [rectangle, rounded corners, minimum width=2cm, minimum height=0.5cm,text centered, draw=black, fill=red!30]
\tikzstyle{process} = [rectangle, minimum width=2cm, minimum height=0.5cm, text centered, draw=black, fill=orange!30, align=left]
\tikzstyle{decision} = [diamond, minimum width=1.0cm, minimum height=0.4cm, text centered, draw=black, fill=green!30]
\tikzstyle{arrow} = [thick,->,>=stealth]

\newcolumntype{L}{>{$}l<{$}}

\usepackage[labelformat=simple]{subcaption}

\usepackage[nolist]{acronym}
\begin{acronym}
\acro{LM}{language model}
\acro{LLM}{large language model}
\acro{SLM}{small language model}
\acro{RAG}{retrieval augmented generation}
\acro{LM}{language model}
\acro{LoRA}{low-rank adaptation}
\acro{QLoRA}{quantized low-rank adaptation}
\acro{SBERT}{sentence BERT}
\acro{ML}{machine learning}
\acro{TF-IDF}{term frequency and inverse document frequency}
\acro{OOD}{out-of-distribution}
\acro{QnA}{questions and answering}
\acro{AI}{artificial intelligence}

\acro{3GPP}{third generation partnership project}
\acro{O-RAN}{open radio access networks}
\acro{MCQ}{multiple choice question}

\acro{MAS}{multi-agent system}
\acro{CRN}{cognitive radio network}

\acro{MDP}{Markov decision process}
\acro{IC}{incentive compatibility}
\acro{IR}{individual rationality}
\acro{URLLC}{ultra reliable low latency}
\acro{AI}{artificial intelligence}
\acro{SNR}{signal-to-noise-ratio}
\acro{CI/CD}{continuous integration and continuous deployment}
\acro{GAN}{generative adversarial network}
\acro{NFV}{network function virtualization}
\acro{SDN}{software-defined networking}
\acro{RLHF}{Reinforcement learning from human feedback}

\end{acronym}

\begin{document}
\captionsetup[figure]{labelfont={bf},labelformat={default},labelsep=period,name={Figure}}





\title{Rethinking Strategic Mechanism Design In The Age Of Large Language Models: New Directions For Communication Systems}



\author{Ismail Lotfi, Nouf Alabbasi and Omar Alhussein

\thanks{
Ismail Lotfi, Nouf Alabbasi and Omar Alhussein are with the Department of Computer Science, Khalifa University, Abu Dhabi, UAE. 

}
}




\maketitle
\begin{abstract}
This paper explores the application of large language models (LLMs) in designing strategic mechanisms —including auctions, contracts, and games— for specific purposes in communication networks. 
Traditionally, strategic mechanism design in telecommunications has relied on human expertise to craft solutions based on game theory, auction theory, and contract theory. 
However, the evolving landscape of telecom networks, characterized by increasing abstraction, emerging use cases, and novel value creation opportunities, calls for more adaptive and efficient approaches. 
We propose leveraging LLMs to automate or semi-automate the process of strategic mechanism design, from intent specification to final formulation. This paradigm shift introduces both semi-automated and fully-automated design pipelines, raising crucial questions about faithfulness to intents, incentive compatibility, algorithmic stability, and the balance between human oversight and artificial intelligence (AI) autonomy. 
The paper discusses potential frameworks, such as retrieval-augmented generation (RAG)-based systems, to implement LLM-driven mechanism design in communication networks contexts. 
We examine key challenges, including LLM limitations in capturing domain-specific constraints, ensuring strategy proofness, and integrating with evolving telecom standards. 
By providing an in-depth analysis of the synergies and tensions between LLMs and strategic mechanism design within the IoT ecosystem, this work aims to stimulate discussion on the future of AI-driven information economic mechanisms in telecommunications and their potential to address complex, dynamic network management scenarios.

\end{abstract}



\section{Introduction}

Several theoretical concepts have been adopted from information economics literature to model, design new architectures, and solve problems in the field of wireless communication and networking
~\cite{Han_Niyato_Saad_Basar_2019, Yanru_2017_MWC_ContractTheory, Yang_2013_COMST_AuctionTheory}.
Game theory is used in communication networks to model and analyze strategic interactions among multiple decision-makers, such as network operators and users~\cite{Han_Niyato_Saad_Basar_2019}. Game theory helps in designing efficient algorithms for resource allocation, spectrum sharing, and power control by predicting and optimizing behaviors in competitive environments. 
For instance, in a setting where two mobile networks operators share the same frequency band, each operator must decide how much power to allocate to their transmissions to maximize their own throughput while minimizing interference to the other. Using game theory, both operators model the scenario as a non-cooperative game, where they adjust their power levels based on each other's actions. The game reaches a Nash equilibrium when neither operator can improve their performance without changing the other's strategy. This approach helps design efficient spectrum sharing algorithms, ensuring both operators use the spectrum optimally while avoiding harmful interference.

While game theory provides a robust framework for understanding and solving complex problems in dynamic and decentralized networks, auction theory and contract theory, as a mechanism design tools, were adopted for different objectives.
Auction theory is used in communication networks to model competitive bidding scenarios, e.g., to induce efficient spectrum resource allocation~\cite{Yang_2013_COMST_AuctionTheory}.
Contract theory helps create incentive-compatible agreements in scenarios with information asymmetry between network operators and users, ensuring optimal resource usage and adherence to service level agreements~\cite{Yanru_2017_MWC_ContractTheory}. Both theories enhance fairness, efficiency, and strategic behavior in telecom systems. The use of these mechanism design frameworks provide theoretically-proven incentives to participants to join the network market and have a mutual-benefit and fair agreements with the participating parties.

The recent rise of \ac{AI}-driven assistant models like ChatGPT, Gemini, and Claude has transformed how individuals interact with \ac{AI} tools, increasingly using them to streamline tasks and enhance productivity in various fields.
For instance, \acp{LLM} are used in chat-bot for customer service automation in online platforms.
Telecom systems are amongst top industries witnessing a remarkable paradigm shift where \ac{AI} and LLMs are heavily adopted to enhance system efficiency. For instance, due to the complexity and large volume of standardization in telecom systems, e.g., \ac{3GPP} standards, \ac{LLM}-based frameworks are proposed recently to solve \ac{MCQ} tasks related to telecom systems~\cite{bornea2024_Telco_RAG}. The early results are promising as near-human level accuracy is achieved when incorporating state-of-the-art \ac{LLM} models. To this end, we envision how generative \ac{AI} tools can be used to enhance existing strategic mechanism design frameworks.

In the telecom industry, an automated LLM-based system that masters senior researchers' expertise, such as the formulation of new or reoccurring problems using appropriate mathematical tools, would bring significantly practical value.
For example, in the context of \acp{CRN}, LLMs can automate the creation of strategic mechanisms for spectrum sharing. For instance, when users dynamically access spectrum, LLMs can help in generating game-theoretic models that optimize spectrum allocation while minimizing interference between primary and secondary users. An LLM-based communication can be initiated automatically between the CRN nodes (i.e., primary and secondary users) to communicate and create the mechanism (a game or an auction) and reach the optimal solution(s).
Without LLMs, human experts would need to manually analyze user demands, interference levels, 
and develop allocation strategies for different scenarios, which is costly and time consuming.

\begin{figure*}[ht!]
    \centering
    \includegraphics[width=.85\textwidth,height=7cm]{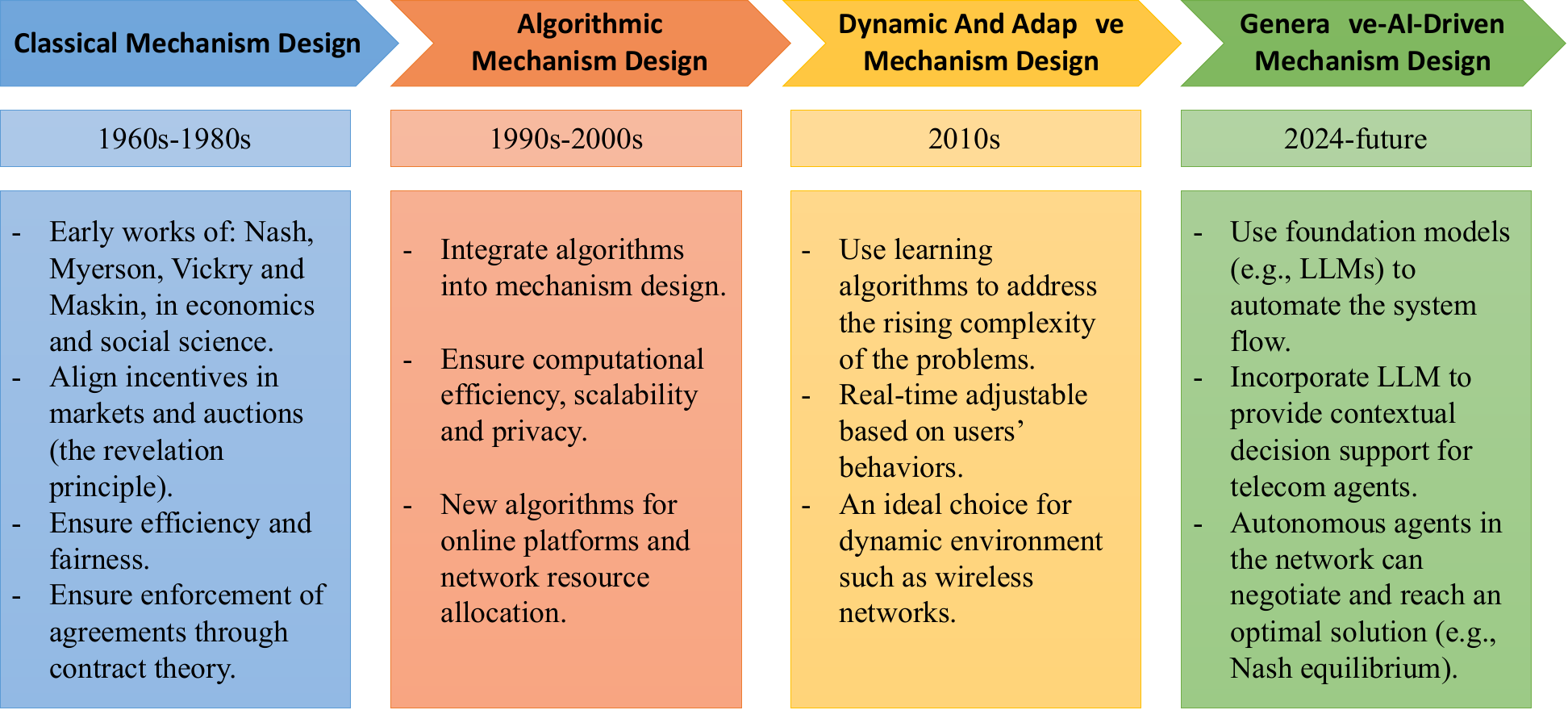}
    \caption{Evolution of strategic mechanism design techniques.}
    \label{fig:evolution_mechanisms}
\end{figure*}

Given the rapid evolution of technology and the increasing integration of LLMs in various fields, there are numerous promising research directions within the context of wireless communication and telecom systems. In this paper, we present our vision for potential research directions that focus on the synergy between generative AI (with a focus on LLMs), contract theory, auction theory, and mechanism design.
Specifically, we present our vision for a new paradigm shift where strategic mechanism design solutions can be automated by leveraging the power of \acp{LLM}.
Automating strategic mechanism design with AI significantly reduces the time needed to model and solve real-time constraints, a task that human experts take longer time periods to address due to the complexity of studying the problem, formulating objectives, and debugging code.
Moreover, telecom standards, such as \ac{3GPP}, evolve rapidly, necessitating adaptable mechanisms that can integrate new technologies and policies. 
Importantly, we believe that the realization of strategic mechanism design automation will be one of the key enablers of zero-touch networks, where minimal human intervention is required.
Therefore, the use of generative AI to explore new strategic mechanism reduces significantly the efforts, otherwise required by human experts.
In this work we attempt to address the following issues:\\

\noindent\textbf{\textit{Q1:} How can we use \acp{LLM} to automate existing strategic mechanism design frameworks?}

\noindent\textbf{\textit{Q2:} To which extent we can minimize human intervention in the design of the strategic mechanisms?}

\noindent\textbf{\textit{Q3:} What are the current challenges and research opportunities to enable generative AI based strategic mechanism design?}

We distinguish our work by the following main novel contributions:
\begin{itemize}
    \item To address \textbf{Q1}, we propose the use of \acp{LM} as a communication enabler between all the network entities (agents) to enable autonomous interactions. As such, based on the network intents, modeling and solving the appropriate strategic mechanism design is handled autonomously through \acp{LM}. 

    \item To address \textbf{Q2}, we explore \emph{semi-automated} strategic mechanism design and \emph{fully-automated} strategic mechanism design frameworks where in the former, a minimal human innervation is required to validate the output of the system, while in the earlier no human intervention is needed. 

    \item To address \textbf{Q3}, we explore key challenges hindering the full integration of \acp{LLM} in strategic mechanism design. Additionally, we outline our perspective on potential strategies to overcome these obstacles. 
\end{itemize}

\section{Preliminaries}

Here, we introduce three foundational theories in economics and decision theory: game theory, auction theory, and contract theory.
Each theory plays a crucial role in understanding and modeling complex economic and social interactions, providing frameworks for analyzing decision-making processes, allocation mechanisms, and contractual agreements in various contexts.
We then provide an overview of the \ac{RAG} framework and show why this framework is suitable for the objective of synergizing generative AI with strategic mechanism design frameworks.

\subsection{Strategic Mechanism Design}

\begin{figure*}[ht!]
    \centering
    \includegraphics[width=.85\textwidth,height=6cm]{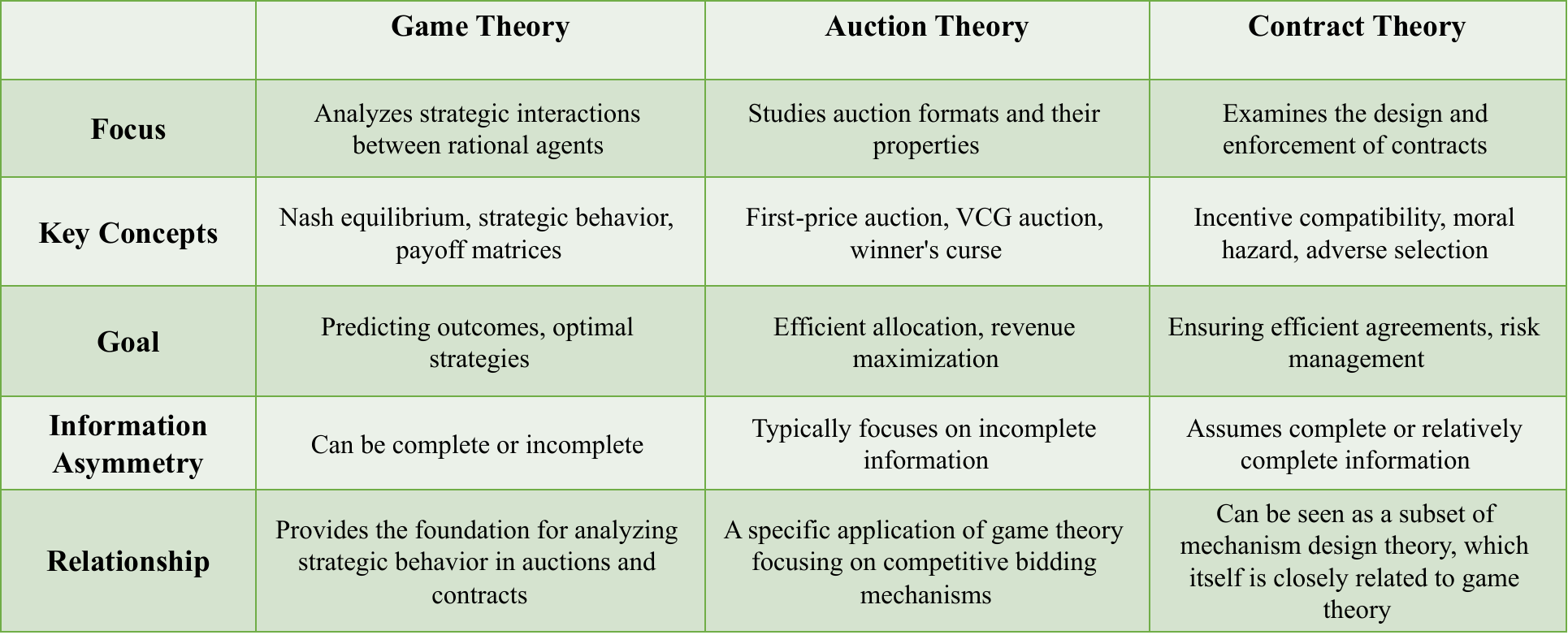}
    \caption{Comparison of different strategic interaction design tools.}
    \label{fig:table_comparision}
\end{figure*}

Figure~\ref{fig:evolution_mechanisms} illustrates the evolution of strategic mechanism design techniques.
The figure captures the progression from classical mechanism design to AI-driven mechanism design, with each phase highlighted by specific use cases.
Each phase of evolution has built upon the previous, incorporating more sophisticated tools and addressing increasingly complex challenges in strategic decision-making and resource allocation.

\subsubsection{Game Theory}
Game theory is a branch of mathematics and economics that analyzes strategic interactions between multiple rational players. It provides a framework for analyzing situations where the outcome of one agent's decision depends on the decisions of others. Key concepts include: Nash equilibrium, payoff matrix, and strategic dominance. 
Different games has been developed by the research community, such as Markov games, Stackelberg games, and zero-sum games.
These game models have been applied to different problems in communication networks such as: sponsored content, resource allocation, power control, and load balancing~\cite{Han_Niyato_Saad_Basar_2019}.
Game models are highly beneficial for network automation as they can effectively model and predict the behavior of network entities in distributed systems. These models allow entities the freedom to plan their moves independently, based on their individual objectives and the incomplete local information available to them. Therefore, by exploring the power of \ac{AI}, we can assist in selecting the optimal game based on information scenarios, such as Nash or Bayesian equilibrium.

\subsubsection{Auction Theory}
This field zooms in on a specific type of games: auctions. Here, multiple players compete for a scarce resource or good by submitting bids. Auction theory studies how to design these auctions to achieve desired outcomes, such as maximizing revenue or ensuring the good goes to the player who values it most. Key concepts include: bidding strategies, auction formats and winner's curse. Several types of auctions have been developed such as: the first-price auctions, the second-price (Vickrey) auctions, ascending (English) auctions and combinatorial auctions.
Similarly, these auction models have been applied to different problems in communication networks such as: spectrum auction, bandwidth allocation and interference management~\cite{Yang_2013_COMST_AuctionTheory}.

\subsubsection{Contract Theory}
Here, the focus shifts to agreements between two parties with conflicting interests or information asymmetries. Contract theory analyzes how to design enforceable agreements that incentivize both parties to act in a way that benefits the overall outcome. It examines how contracts can align incentives, manage risks, and facilitate efficient transactions.
Contract theory involves designing mechanisms for negotiation and contract formation, ensuring that agreements are enforceable and efficient. Key concepts include: incentive compatibility, adverse selection and moral hazard.
Applications of contract theory to communication networks problems include spectrum sharing in \acp{CRN}, quality of service (QoS) agreement and cooperative communication~\cite{Yanru_2017_MWC_ContractTheory}.

To sum up, auction theory and contract theory borrow heavily from game theory to analyze specific scenarios where strategic decision-making is crucial for achieving optimal outcomes. Auction theory focuses on situations where competition determines who gets what, while contract theory focuses on agreements between two parties that specify rights and obligations. A detailed comparative overview of game theory, auction theory and contract theory is shown in Figure~\ref{fig:table_comparision}.

\subsection{Retrieval Augmented Generation}

A critical limitation for off-the-shelf \acp{LM} is their inability  to tune their knowledge beyond their training dataset, which can cause the LM to output outdated information. Additionally, the off-the-shelf LM does not have external data source to rely on for in-depth answers. To address these limitations, \ac{RAG} was proposed and is gaining a considerable interest by telecom experts~\cite{zou2024_TelecomGPT}. RAG is useful for providing up-to date information, which can minimize the need to retrain the whole LM model for new set of information. For instance, if a new strategic mechanism design is newly proposed by the research community, it can be easily integrated as an additional context throughout RAGs. From the other side, we might want to refrain from using certain mechanisms as they become deprecated for some newly apparent drawbacks.  

One of the main advantages of using RAGs in telecom is the rapidly changing environment of the LLM agents. 
As the LM relies only on the internal reasoning of its model, it cannot make new optimal strategies or decisions without relying on new information about the environment, and here is where RAGs become crucial. 
Incorporating retrieved information into the prompt enables the model to respond in an “open-book” setting, using external documents to inform its answers instead of relying solely on its internal knowledge.
For example, new context about energy consumption of an IoT device and \ac{SNR} are needed for the agent to make new decision about their transmit power. The new context is then fed to the LLM through RAG.

\begin{figure*}[ht!]
    \centering
    \includegraphics[width=.85\textwidth,height=8cm]{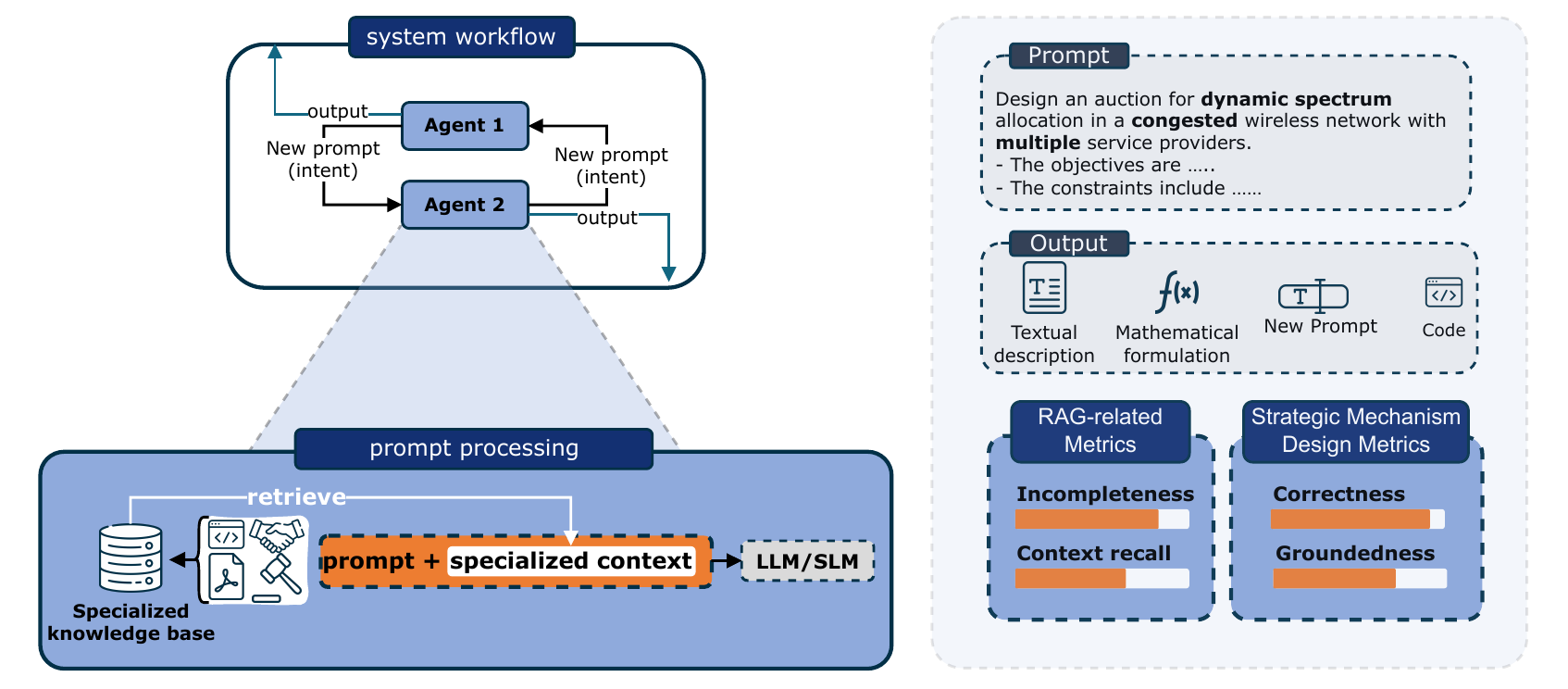}
    \caption{Automated strategic mechanism design with prompt-based communication.}
    \label{fig:automatic_mechanism}
\end{figure*}

\section{Synergizing LLMs With Strategic Mechanism Design}

The main objective for integrating LLMs in the design of strategic mechanisms is to minimize the human intervention required.
In what follows, we first start by presenting our vision of how such integration can become feasible and present the challenges that need to be addressed to achieve strategic mechanism design automation for communication networks.
We also provide some use cases to further motivate the need for such integration and show how this synergy can enable automated telecommunication systems.

\subsection{General Workflow}



Figure~\ref{fig:automatic_mechanism} depicts the workflow of a prompt received by an LLM agent (e.g., mobile operator) and the processing phases it goes through until the final output is delivered. 
Specifically, once the prompt is received by an agent, it gets augmented using a specialized context retrieved from the knowledge base documents (books, scientific papers and code). 
Here, the knowledge base can either contain all documents related to strategic mechanism design frameworks (i.e., game theory, auction theory and contract theory) or a portion of the documents. 
In case where all related documents are fetched, the retrieved context is directly forwarded to the LLM or the \ac{SLM} next to the original prompt. Otherwise, a two-step architecture can be implemented where the system first decides which specific framework will be used to fetch the relevant chunks from then the prompt is forwarded to that specialized knowledge base to retrieve the relevant pieces of information.
Finally, the LLM/SLM output a textual description of the strategic mechanism design with all necessary mathematical formulation and code. Additionally, a new prompt is generated, if necessary, for the other agents to indicate any required additional actions.

To ensure the system's output meets quality standards, an additional validation step can be introduced in which feedback scores are gathered and used to refine the model's performance. This feedback process can involves human evaluators assessing the outputs based on specific criteria and providing scores that are then fed back into the language model. This additional feedback loop allows the model to improve its accuracy, relevance, and reliability iteratively. 
As illustrated in Figure~\ref{fig:automatic_mechanism}, the feedbacks received on the LM output can be classified into two categories: feedbacks related to the quality of the formulated strategic mechanism and feedbacks related to the quality of the RAG architecture.

\textit{1) Strategic mechanism design metrics:}

\textbf{Correctness:} A response is considered correct only when it is both factually accurate and relevant to the query intent, while the presence of any factual inaccuracy renders the entire response incorrect. For instance, a response to an auction query that discusses an equilibrium is considered incorrect as such discussions are only present in game theoretic settings.

\textbf{Groundedness:} A correct response's groundedness is determined by whether its key factual claims are properly supported by relevant technical sources, with ungrounded responses lacking citations and misgrounded ones either misinterpreting sources or citing irrelevant ones. For instance, the retrieved text from technical documents might suggest the utility function must be convex as the objective function from the retrieved documents was a maximization problem. However, in the current problem, the objective might be to minimize the objective and therefore the utility function should be crafted to be concave. Whether the LLM can infer such relationship is captured by the groundedness metric.

\textit{2) RAG-related metrics:}

\textbf{Incompleteness:} A response is considered incomplete when it fails to address all the relevant aspects or requirements of the query, omitting key information that was explicitly requested or implicitly needed for a comprehensive response. This differs from incorrectness or hallucination as the information provided may be accurate but insufficient for the full scope of the question. For instance, a response to a game theory query that does not show the existence of the equilibrium is considered incomplete.

\textbf{Context recall:} This metric refers to how well the model remembers and appropriately uses the relevant information provided in its input context or prompt. This differs from general recall (completeness) as context recall specifically focuses on utilizing the information provided in the immediate prompt/context, not the model's broader knowledge.

Once collected, these metrics can be then incorporated into an RLHF framework to enhance the LM output in future rounds, e.g., as in~\cite{wu2024_RHLF_NIPS}.

\subsection{Use Cases}

\begin{figure*}[ht!]
    \centering
    \includegraphics[width=.85\textwidth,]{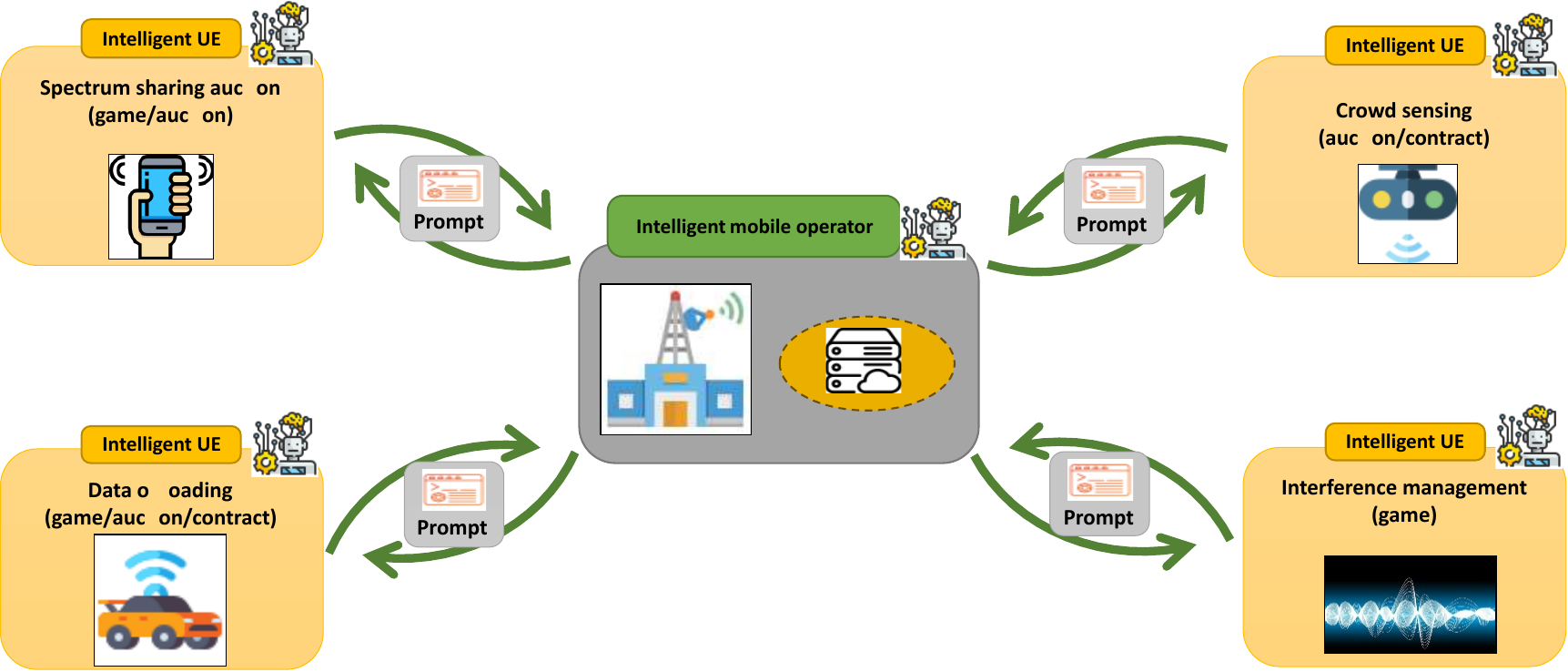}
    \caption{Different use cases in communication systems for strategic mechanism design automation.}
    \label{fig:exemplary_use_cases}
\end{figure*}

Figure~\ref{fig:exemplary_use_cases} highlights different use cases of the envisioned framework.
The framework's choice between the three strategic mechanism design theories is based on the specific nature of the communication network issue being addressed.
If the problem involves strategic interactions and competition among multiple agents, game theory is applied. Game theory's ability to model complex interdependencies between multiple entities makes it essential in achieving stable network configurations. For instance, in scenarios like resource allocation and spectrum sharing, where multiple network users must minimize interference (e.g., by reducing transmit power levels), game theory predicts behavior to achieve 
equilibrium. 

In cases where resource allocation is based on bids, auction theory is employed. Auction theory requires detailed knowledge of bidding behaviors and market dynamics. A typical example is allocating spectrum bands to IoT devices, where the system designs auction mechanisms that are efficient and fair, which is vital in competitive markets like telecommunications. 
Contract theory, on the other hand, becomes the primary focus when the challenge involves designing incentive structures in environments with asymmetric information. The system leverages this framework to ensure cooperation between parties, such as network providers and users, by aligning incentives. This is crucial in scenarios like data offloading during peak times, where the system must design contracts that benefit both users and providers.

\subsection{Automated Strategic Mechanism Design}

Through iterative cycles of the proposed prompt-based communication approach, agents within the network can autonomously interact and collaboratively solve problems. This process establishes a new paradigm for automated strategic mechanism design driven by prompt-based communication.
The emergence of using LLMs as optimizers is key for this autonomous problem-solving design~\cite{yang2024_LLMs_as_optimz}.
However, a critical element of strategic mechanism design is the satisfaction of the derived solutions to certain formal properties~\cite{Han_Niyato_Saad_Basar_2019}. For example, in zero-sum games, the Nash equilibrium should be proven to exist, while in auction theory, incentive compatibility should be proven to be satisfied for all bidders. Here, we envision that the integration of generative AI into strategic mechanism design can be realized in two different directions, which we done as \textbf{``semi-automated strategic mechanism design"} and \textbf{``fully-automated strategic mechanism design"}.

\subsubsection{Semi-Automated Strategic Mechanism Design}
In this approach, the human validation is required after the generation of every new strategic mechanism (e.g., a game or an auction). This requirement is due to the hallucination problem of LLMs~\cite{huang2023_survey_hallucination_LLMs}, where the model might generate seemingly accurate proofs that are, in fact, a mix of contradictory and nuanced statements. Therefore, an expert's validation is required before deploying such solution to prevent system failure. For example, in the framework of game theory, if the Nash equilibrium is not valid, the adversary player can exploit the system and have a high advantage of maximizing their profit at the cost of reduced profit of the system designer. 
Nevertheless, this semi-automated approach faces some critical challenges:

\begin{itemize}
    \item First, the human validation is not open to any person with general knowledge such as that in image classification tasks. Indeed, only experts within the field of communication networks and application of strategic mechanism design frameworks can perform the validation tasks. 
    Such human experts are a scarce resource which puts more burden on the entity willing to deploy such framework, even for large telecom companies.

    \item An implicit limitation of the above challenge is that the framework becomes practical only with non-real-time settings. In other words, as human validation is a mandatory step in the semi-automation process, the framework can be used only in situations where immediate responses are not required. This delays decision-making and affects real-time applications, particularly those that demand rapid processing, such as dynamic network optimization or on-the-fly spectrum allocation. This constraint makes the framework unsuitable for environments demanding quick responses, such as multi-agent systems where real-time communication and coordination between multiple autonomous entities are essential. 
    
\end{itemize}

Although the aforementioned limitations are critical, the framework still reduces significantly the human intervention from the end-to-end system perspective. The classical and current approach requires the human expert to 1) find the appropriate framework to use then, 2) formulate the problem using that framework then, 3) derive the necessary proofs and then, 4) write the code. However, in the proposed semi-automated approach, the human intervention is limited to the validation step.

Next, we present an enhanced version of the semi-automated framework, in which we show how the limitations of the semi-automated framework could be addressed.

\subsubsection{Fully-Automated Strategic Mechanism Design}

\begin{figure}[ht!]
    \centering
    \includegraphics[width=.40\textwidth,]{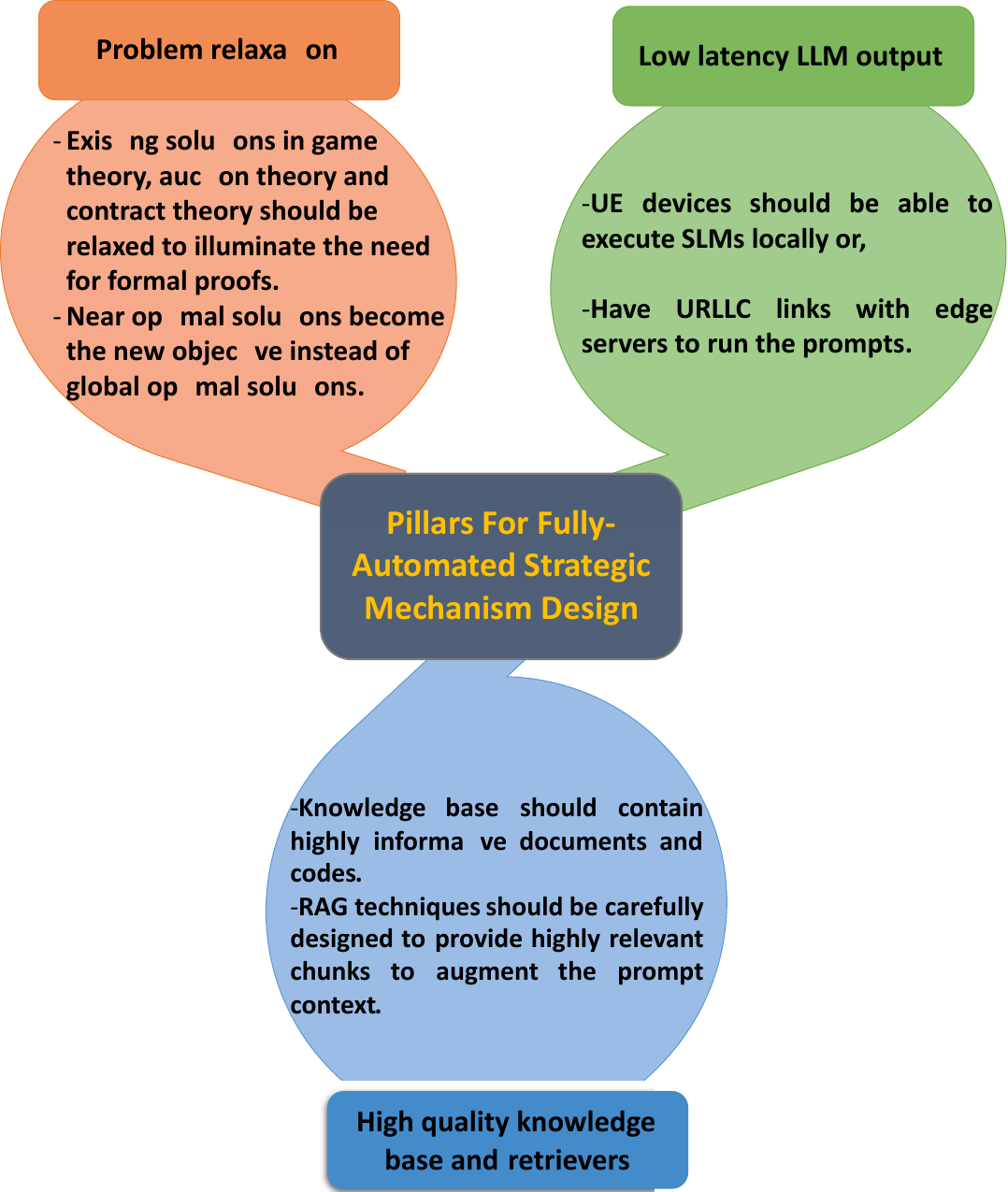}
    \caption{Three essential pillars for fully-automated strategic mechanism design.}
    \label{fig:pillars_automated_mechanism}
\end{figure}

To achieve a fully automated strategic mechanism design, we highlight three key criteria to be addressed as depicted in Figure~\ref{fig:pillars_automated_mechanism}. First, existing theoretical solutions of strategic mechanism design problems should be relaxed, favoring near-optimal outcomes over global optimality, reducing the need for formal proofs. Second, devices must ensure low-latency by either running \acp{SLM} locally or using \ac{URLLC} links to edge servers. Lastly, high-quality knowledge bases and well-designed \acp{RAG} systems are crucial for providing relevant information to enhance prompt responses and system accuracy.

Importantly, the frameworks of game theory, auction theory, and contract theory can be restructured towards relaxed objectives. Instead of setting strict conditions for the strategic mechanism to be acceptable, the new paradigm will allow certain constraints to be violated occasionally subject to the global objective of the system (usually defined as the payoff function) is maximized. 
Though limited, recent studies are beginning to explore this direction, such as the works found in~\cite{Dutting_2024_LLM_Auction, Ismail_2023_JSAC}. For instance, Ismail et al. suggest a new iterative contract model in which the contract problem is framed using a \ac{MDP} and the objective is to maximize a long-term reward function that integrates both the contract objective and constraints~\cite{Ismail_2023_JSAC}. Compared to the standard framework of contract theory where the \ac{IC} and \ac{IR} properties are required to be satisfied for all the participants, a relaxed objective is formulated within the reward function of the \ac{MDP} in~\cite{Ismail_2023_JSAC}. Specifically, the reward function is crafted as a weighted sum of the contract designer payoff and the number of \ac{IC} and \ac{IR} violations. Such a relaxed system can achieve a near-optimal solution where the system designer's payoff is maximized and the IC and IR violations are minimized. 
Importantly, in such systems, no mathematical proof is required, which is a major limitation for the semi-automated approach.
Therefore, we argue that if more relaxed forms of strategic mechanism design are available, we can use that as part of the knowledge base of the specialized-RAG module as shown in Figure~\ref{fig:automatic_mechanism}. As such, the human intervention for proof validation becomes unnecessary. 

Another critical requirement to enable the full automation of strategic mechanism design is the latency for receiving the LLM output, which has to be minimized. To achieve a low latency LLM output, agents with low computing power should be able to execute language models locally (e.g., \acp{SLM}) or, have a \ac{URLLC} link with edge servers to offload prompts execution. Achieving a low latency LLM output is crucial for the system as interaction between autonomous agents in multi-agent systems requires real-time decisions.

To sum up here, we believe a temporary step back is essential at this juncture. Human expertise remains crucial for devising novel solutions and architectures in strategic mechanism design. Once these innovations are developed, they can be leveraged by LLMs, paving the way for full automation in this domain. This phased approach ensures that foundational advancements led by human intelligence will eventually empower AI to manage complex, automated processes with minimal oversight.

\section{Future Directions, Open Issues And Potential Impact}

\subsection{Historical Strategy Memorization}
To achieve the oracle of automated strategic mechanism design, agents in the system should be able to perceive their current state and then self-generate prompts to communicate with other agents. 
Furthermore, agents need to memorize previous histories from different games, auctions and contracts. 
However, using a Markovian system would cause the LLM to lack the contextual understanding required to optimize for long-term goals. Recent research such as in~\cite{yang2024_LLMs_as_optimz} highlights this limitation, showing that LLMs would be unable to adapt effectively without considering the task’s broader context and past interactions.
The framework of repeated games theory could be a key enabler to model agents' states across timesteps~\cite{Hoang_2015_COMST_RepeatedGames}.
Repeated games theory is crucial in modeling agents' states over time, as it accounts for past interactions unlike \acp{MDP}, which only focus on the present.

The new challenge in historical strategy memorization arises because traditional frameworks, like repeated games theory, model strategies within the same game types and strategies. However, in dynamic systems involving varied games, auctions, and contracts, LLMs must optimize future strategies across different modalities, which extends beyond repeated games' scope. LLMs introduce new possibilities by offering the potential to self-generate prompts and learn from diverse interactions. However, their limitation lies in lacking contextual depth from past actions, which makes adapting across different game types a novel area for research and optimization.

\subsection{Strategic Mechanism Design Validation}
User satisfaction with LLM-generated outputs can serve as a critical feedback mechanism to improve system performance. While a system may generate technically feasible and correct strategic mechanisms, these may not align with the actual goals or intents of users. To bridge this gap, \ac{RLHF} can be employed to refine the model iteratively. RLHF enables the model to learn from user preferences and adapt its outputs, ensuring that the generated solutions not only meet formal requirements but also capture user intent more accurately~\cite{wu2024_RHLF_NIPS}. This continual learning process, driven by human feedback, ensures that the system evolves and becomes more effective over time, ultimately enhancing both system accuracy and user satisfaction. 
Additionally, using RLHF fosters dynamic interaction between the LLM and users, improving adaptability, especially in evolving strategic environments like telecommunications, where preferences and conditions change frequently. Over time, the system can fine-tune its understanding of users’ goals and increase its robustness across different strategic scenarios.

The power of generative AI can be further used to validate the correctness of the provided solution in both semi-automated and fully-automated strategic mechanism design approaches through digital twin. 
Exploring the use of digital twin is strategic mechanism design is expected to bring valuable benefit to the overall system performance.
For human validation, digital twin can be used to ensure solution accuracy. 
For instance, \acp{GAN} can simulate various network conditions, such as user density and interference. LLMs generate game or contract strategies, and the GAN’s generator-refines the model while the discriminator evaluates against realistic conditions. Introducing adversarial conditions tests the robustness of strategies, ensuring resilience to unexpected behaviors.

\subsection{Reasoning In Strategic Mechanism Design Proofs}
In game theory, the proof of the existence of the equilibrium relies heavily on the definition of the payoff function and the payoff function needs to be well justified. 
For instance, a simple convex payoff function can simplify the proof process but might have irrelevant physical interpretation to the current communication networking problem.
This issue raises concerns about LLMs’ capability to manage both tasks of modeling the problem and providing the corresponding proof accurately.
Furthermore, LLMs often experience hallucinations when they fail to capture the dynamic variations in rapidly changing networks, leading to incorrect conclusions. Hallucinations can also occur due to inter-agent interference or inadequate data. To reduce such errors, LLMs need better reasoning tools, such as causal inference frameworks, and architecture modifications that enable logical proofs.

\subsection{Hybrid Strategic Mechanism Design}
Recent studies suggest that a superior efficiency, fairness, and adaptability can be achieved when these theories (game theory, auction theory and contract theory) are applied together~\cite{Huang2013, Wang_Yu_2024} . For instance, Huang et al. demonstrated that their hybrid and human-designed mechanism is able to provide superior surplus compared to the standard VCG mechanism~\cite{Huang2013}. 
Therefore, instead of using a different retrieval for each theoretical framework as suggested in our automated framework, a hybrid but carefully designed retrieval is expected to lead to new insightful and promising results beyond the separate use of these theoretical frameworks independently.

Furthermore, Mix-LoRA technique can be used for efficient fine-tuning of the \ac{LLM}/\ac{SLM} by mixing low-rank adapters, allowing the model to adapt to different types of strategic problems (e.g., auctions vs. contracts) without massive computational overhead~\cite{li2024_mixlora}. 
Additionally, the use of mixture of experts technique which dynamically routes different parts of the model for specific tasks, allows efficient scaling and specialization, which can improve decision-making for various strategic mechanisms.
Together, these approaches are expected to offer scalability and flexibility for automating strategic mechanism design and are worth exploring.

\subsection{Impact Beyond Communication Networking Systems}

Disciplines such as economics and social sciences, where strategic mechanism design initially originated, can benefit from the solutions developed here with minimal adjustments. Specifically, the proposed framework can be adapted to other domains by simply altering the knowledge base content to reflect the specific documents and terminology of that field. This flexibility allows the system to extend its applicability beyond telecommunications, making it versatile for various application scenarios while maintaining its core functional design.

\section{Conclusion}
The integration of generative AI into strategic mechanism design for telecom systems presents transformative opportunities for reducing human intervention and improving decision-making efficiency. 
Through the proposed frameworks, including semi- and fully-automated approaches, AI systems can autonomously design and optimize communication networking strategies.
These advancements promise to address real-time constraints, adapt to evolving telecom standards, and deliver efficient resource allocation. However, significant challenges remain, such as enhancing AI reasoning capabilities, managing resource constraints, and ensuring system robustness. 
Future research must focus on overcoming obstacles like AI model vulnerabilities, heterogeneous system coordination, and improving reasoning to reduce hallucinations. Additionally, attention must be given to securing AI models against prompt injection attacks and enhancing collaboration between varied network agents.

\bibliographystyle{IEEEtran}
\bibliography{ref}

\end{document}